\documentclass[12pt]{article}
\usepackage{amsmath,amssymb,amsthm}
\usepackage{graphicx}
\usepackage{natbib}
\usepackage{setspace}
\usepackage{geometry}
\usepackage{booktabs}
\usepackage{bm}
\usepackage{hyperref}
\title{
Mutation Rate Variation Across Genomic Regions in \textit{Arabidopsis thaliana}}

\author{
Elisa Heinrich-Mora$^{1}$ \and Marcus W. Feldman$^{1}$\thanks{Corresponding author: Marcus W. Feldman, \href{mailto:marcuswf@stanford.edu}{mfeldman@stanford.edu}}
}

\date{
\small
$^{1}$Department of Biology, Stanford University, Stanford, CA, USA
}

\begin{document} 

\maketitle 
\begin{abstract}
In population genetics, mutation rate is often treated as a homogeneous parameter across the genome. Empirical evidence, however, shows systematic variation across genomic contexts associated with chromatin organization and epigenomic features. Using gene-level de novo mutation data from \textit{Arabidopsis thaliana}, we test whether chromatin features predict not only the mean per–base mutation rate but also its variability across genes. To reduce heterogeneity in selective regime, we restrict analysis to essential and lethal loci subject to strong purifying selection. Across complementary multivariable models—including heteroskedasticity-robust linear regression, length-weighted regression, and Poisson generalized linear models with exposure offsets—histone marks associated with active transcription (H3K4me1, H3K4me3, H3K36ac) are consistently associated with lower mean mutation rates and substantially reduced between-gene variance. GC content shows little association with the mean once chromatin predictors are controlled but is positively associated with mutation-rate variability. Estimates of skewness and kurtosis reveal no significant higher-order structure attributable to epigenomic predictors. A standardized Tajima’s $D$ statistic yields directionally consistent but statistically underpowered associations with both the mean and variance of gene-level mutation rates. These results suggest that mutation rate is systematically structured by chromatin state within functionally constrained genes and suggest that evolutionary processes may act not only on expected mutation rate but also on its variability across loci.
\end{abstract}

\noindent\textbf{Keywords:} mutation rate; chromatin; epigenomics; \textit{Arabidopsis thaliana}; population genetics

\section{Introduction}
In population genetics, mutation rate is often assumed to be homogeneous across the genome, acting as a constant input into evolutionary processes. Mutations are assumed to arise independently at a constant rate $\mu$ per site per generation, so that the number of new mutations in a genomic region of length $L$ follows a Poisson distribution with mean $\mu L$. This abstraction has been useful, supporting formal analysis of mutation–selection balance, genetic load, and molecular evolution \cite{haldaneEffectVariationFitness1937,fisherXVIITheDistributionGene1931,ewensMathematicalPopulationGenetics2004}. The assumption of independence and homogeneity, however, is methodological rather than empirically validated.

Empirical work over the past two decades shows that mutation rates vary across genomic contexts. Mutation frequency differs among chromatin states and correlates with transcriptional activity, DNA methylation, replication timing, and local sequence context \cite{schuster-bocklerChromatinOrganizationMajor2012,makovaEffectsChromatinOrganization2015,monroe_mutation_2022}. In \textit{Arabidopsis thaliana}, direct estimates of de novo mutations indicate that gene bodies marked by histone modifications associated with active transcription have lower mutation rates than unmarked regions \cite{monroe_mutation_2022}. Comparable patterns have been reported in other eukaryotes and, in modified form, in prokaryotes \cite{martincorenaEvidenceNonrandomMutation2012,polakCelloforiginChromatinOrganization2015,smithLargeScaleVariation2018}. Mutation rate is therefore not homogeneous across the genome, it is influenced by molecular features of chromatin organization, DNA repair activity, and nuclear structure. 

This observation raises two related questions. First, what determines the \emph{mean} mutation rate across genes? Second, what determines its \emph{variability}? Evolutionary theory has focused primarily on the first. The genomic mean mutation rate influences equilibrium load and long-term adaptability, and models predict that selection, drift, and physiological constraints jointly determine its value \cite{leighNaturalSelectionMutability1970,sniegowskiEvolutionMutationRates2000}.

The second question has received less attention. Two genomes may share the same mean mutation rate while differing in how mutational probability is distributed across genomic regions. If a subset of genes experiences episodic or persistently elevated mutation rates, and if those genes are functionally constrained, the fitness consequences may exceed those expected under a uniform mutation rate. Under such conditions, selection might act not only on the mean mutation rate but also on its variance across loci.

This possibility does not require fine-tuned adaptation at each gene. It requires only that genomic features influencing mutation-rate variability be heritable and that increased variance elevate the probability of deleterious mutations in essential regions. Modifiers of mutation rates that reduce extreme excursions in mutation rate, even without altering the mean, could then be indirectly favored.

To examine these ideas empirically, it is useful to focus on a set of genes that are consistently under strong purifying selection. Essential and lethal genes provide such a context. By definition, these loci tolerate little functional disruption, and restricting analysis to this class reduces heterogeneity in selective regime while limiting the possibility that differences in mutation rate simply reflect relaxed constraint in dispensable genes. Within this constrained set, variation in mutation rate is more plausibly attributable to intrinsic genomic mechanisms — chromatin organization, transcription-coupled repair, and local DNA accessibility — than by functional differences subject to differential selection.

Using gene-level de novo mutation data from \textit{Arabidopsis thaliana} \cite{monroe_mutation_2022}, we address three questions. First, are chromatin-associated epigenomic features associated with lower mean per-base mutation rates in essential and lethal genes? Second, are these features associated with reduced variance or altered higher moments of the mutation-rate distribution across genes? Third, does standardized Tajima’s $D$, used here as a population-genetic measure of relative selective constraint, covary with these distributional properties after accounting for chromatin-state predictors?

Tajima’s $D$ compares nucleotide diversity with the number of segregating sites and is sensitive to both selection and demography. Within a restricted functional class, more negative values may reflect stronger purifying selection, although this interpretation is not definitive. We therefore use it as an imperfect but informative proxy for relative constraint.

We estimate partial statistical associations under multiple modeling frameworks. Our aim is not to establish direct causation, but to test whether mutation rate — and in particular its variability — is systematically structured by epigenomic context, in which case, it should exhibit coordinated shifts in both mean and variability within functionally constrained regions. Such structure would indicate that mutation rate is partly determined by the organization of the genome in which it arises.

\section{Data and Statistical Framework} \label{sec:data_methods}
Our objective is to estimate how gene-level epigenomic features covary with mutation rate within a set of functionally constrained genes, while making all statistical assumptions explicit. Three features of the data shape the analysis. First, de novo mutations are rare events, so gene-level counts are sparse. Second, gene lengths vary by orders of magnitude, so raw counts are not directly comparable. Third, chromatin marks and related epigenomic variables are correlated, as they reflect overlapping aspects of nuclear organization. The framework is designed to address these constraints directly. We define a response that is interpretable in per-base terms, we incorporate gene length formally (either through normalization or exposure offsets), and estimate partial rather than marginal associations in multivariable models. We also compare complementary specifications to assess robustness to distributional and weighting assumptions.

\subsection{Gene Set and Data Sources} \label{subsec:geneset_data}
We analyzed genes annotated as essential or lethal in \textit{Arabidopsis thaliana}. These loci are expected to experience consistently strong purifying selection, reducing heterogeneity in selective regimes across genes. De novo mutation counts and gene-level epigenomic annotations were obtained from Monroe et al.\ \cite{monroe_mutation_2022}, who quantified genome-wide mutation-rate variation and its association with chromatin features. We used the gene-level predictors reported in that study (e.g.\ histone modification signal, methylation summaries, accessibility measures).

\subsection{Response Variable: Per--Base Mutation Rate}\label{subsec:response}
For each gene $i = 1,\dots,n$, let $C_i$ denote the number of observed de novo mutations assigned to that gene and let $L_i$ denote its callable coding-sequence (CDS) length in base pairs. We define the empirical per-base mutation rate as
\begin{equation}
\mu_i \;=\; \frac{C_i}{L_i}.
\label{eq:rate_def}
\end{equation}
Under a model in which each base has equal and independent mutation rate, $\mathbb{E}[C_i] = \lambda_i L_i$, so expected counts scale linearly with length. The normalization in Eq.~\eqref{eq:rate_def} therefore yields a quantity directly comparable across genes of different sizes.

However, when $C_i$ is small (often 0 or 1), $\mu_i$ is a noisy estimator whose sampling variance depends strongly on $L_i$. If $C_i \sim \text{Poisson}(\lambda_i L_i)$, then
\[
\mathrm{Var}(\mu_i) = \frac{\lambda_i}{L_i},
\]
which decreases with gene length. Consequently, $\mu_i$ is not homoskedastic across genes. This motivates models that either account for length-dependent variance or treat $C_i$ directly as a count outcome.

\subsection{Scaling and interpretability}
In \citep{monroe_mutation_2022}, some predictors were reported on a 0--100 scale representing percentages of coverage or signal. These are proportions multiplied by 100 and have no intrinsic unit beyond that scaling. We rescaled them to $[0,1]$, removing the arbitrary factor of 100 without changing relative differences among genes.

Because predictors are measured in heterogeneous units, we standardized each variable:
\begin{equation}
X_{ij}^{\ast} \;=\; \frac{X_{ij}-\bar X_j}{\mathrm{SD}(X_j)},
\label{eq:zscore}
\end{equation}
where $X_{ij}$ is the value of predictor $j$ for gene $i$, $\bar X_j$ is its sample mean, and $\mathrm{SD}(X_j)$ its sample standard deviation. Each standardized predictor therefore has mean 0 and variance 1. Regression coefficients represent the expected change in the response associated with a one--standard--deviation increase in the predictor, conditional on the other covariates. Because many chromatin marks are correlated, coefficients estimate partial associations rather than marginal effects.

\subsection{Predictors: Sequence Composition and Epigenomic Context}  \label{subsec:predictors}
We model predictors representing two mechanistically distinct but correlated sources of mutation-rate heterogeneity: local sequence composition and epigenomic organization. Following \cite{monroe_mutation_2022}, these variables are defined at the gene or genic sub-region level and interpreted as features of chromatin state and gene-body architecture that may influence DNA damage and repair efficiency.

\paragraph{Sequence composition.} Local nucleotide composition directly affects mutability through known biochemical mechanisms and also correlates with chromatin organization. In particular, methylated cytosines in CG context are prone to spontaneous deamination, increasing C$\to$T transition rates. We include:
\begin{itemize}
    \item GC content (fraction of G and C bases within the focal region),
    \item CG methylation level (mean fractional methylation at CG sites within the region).
\end{itemize}
These variables capture sequence-level mutability as well as correlated aspects of chromatin state.

\paragraph{Chromatin state and gene-body architecture.}
Mutation rates vary systematically across genic regions and chromatin states. In \cite{monroe_mutation_2022}, mutation frequency was found to be lower in gene bodies of actively transcribed genes and enriched in flanking regions, consistent with preferential DNA repair targeting essential genes. To capture this structure, we include:
\begin{itemize}
    \item Chromatin accessibility (ATAC--seq signal),
    \item Histone modification levels associated with active gene bodies
    (e.g.\ H3K4me1, H3K4me3, H3K36me3, and H3K36ac),
\end{itemize}
summarized over gene bodies or defined genic sub-regions. These predictors are correlated, reflecting overlapping dimensions of nuclear organization. Accordingly, effects are estimated jointly in multivariable models to recover partial associations between each feature and mutation rate.

\subsection{Modeling Strategy: Three Specifications}\label{subsec:models}
Estimated associations can depend on modeling assumptions, including the scale of the response and the treatment of gene length. To evaluate robustness to assumptions about the response scale, variance structure, and treatment of gene length, we fit three model classes (i) linear models for normalized per-base rates, (ii) heteroskedastic linear models that allow variance to depend on gene length, and (iii) generalized linear models for mutation counts with gene length included as an exposure offset. Agreement across these specifications indicates that estimated associations are not driven by a particular distributional or weighting assumption.

\subsubsection{Model 1: OLS with heteroskedasticity-robust inference}\label{subsubsec:ols}
We first model the per-base mutation rate $\mu_i$ directly:
\begin{equation}
\mu_i \;=\; \beta_0 \;+\; \bm{x}_i^{\ast\top}\bm{\beta} \;+\; \varepsilon_i,
\label{eq:ols}
\end{equation}
where $\bm{x}_i^{\ast}$ is the vector of standardized predictors for gene $i$, $\bm{\beta}$ is the coefficient vector, and $\varepsilon_i$ is the residual.

Ordinary least squares (OLS) estimates $\hat{\bm{\beta}}$ by minimizing the sum of squared residuals. Each coefficient $\beta_j$ therefore represents the expected change in per-base mutation rate associated with a one–standard–deviation increase in predictor $j$, holding all other predictors constant. Classical OLS inference assumes \emph{homoskedasticity}, meaning that the residual variance $\mathrm{Var}(\varepsilon_i)$ is constant across genes. In this context, that assumption is unlikely to hold. Mutation counts are sparse, and per-base rates estimated for short genes are intrinsically noisier than those estimated for long genes. Thus, the residual variance plausibly depends on gene length $L_i$, generating \emph{heteroskedasticity}. If heteroskedasticity is ignored, conventional standard errors are typically biased downward, leading to overly narrow confidence intervals and anti-conservative $p$-values.

To avoid specifying a parametric model for how variance scales with gene length, we use heteroskedasticity-consistent (``robust'') standard errors. These estimators leave the OLS coefficient estimates $\hat{\bm{\beta}}$ unchanged but replace the classical variance formula with one that uses the observed squared residuals to estimate $\mathrm{Var}(\hat{\bm{\beta}})$. Under mild regularity conditions, this yields asymptotically valid inference even when residual variance differs across genes. Specifically, we use the HC3 estimator. Among robust estimators, HC3 is designed to perform well in finite samples by accounting for \emph{leverage}. The leverage of observation $i$, denoted $h_i$, measures how unusual its predictor values are relative to the rest of the data. Formally, $h_i$ is the $i$th diagonal element of the projection (``hat'') matrix and quantifies the influence of gene $i$ on its own fitted value. Genes with extreme predictor combinations have high leverage and can exert disproportionate influence on coefficient uncertainty.

HC3 adjusts for this by inflating each squared residual in the variance calculation by a factor $(1 - h_i)^{-2}$. Because $h_i$ increases with leverage, this correction increases the contribution of high-leverage observations to the estimated variance--covariance matrix. The result is typically more conservative and more reliable standard errors in samples where influential observations are present. Importantly, HC3 modifies only the estimated variance--covariance matrix of $\hat{\bm{\beta}}$; the point estimates themselves remain identical to those obtained by OLS. All reported confidence intervals and hypothesis tests are therefore robust to unequal residual variance and to differences in influence among genes.

\subsubsection{Model 2: Length-weighted least squares} \label{subsubsec:wls}
In the OLS specification, coefficients are chosen to minimize the unweighted sum of squared residuals,
\[
\sum_{i=1}^{n} (\mu_i - \beta_0 - \bm{x}_i^{\ast\top}\bm{\beta})^2.
\]
This treats each gene as equally informative about $\bm{\beta}$.

If $C_i \sim \text{Poisson}(\lambda_i L_i)$, then
\[
\mathrm{Var}(\mu_i) = \mathrm{Var}\!\left(\frac{C_i}{L_i}\right)
= \frac{\lambda_i}{L_i},
\]
so the sampling variance of $\mu_i$ is approximately proportional to $1/L_i$. Observations with larger $L_i$ therefore have smaller variance and provide more precise information about the underlying mutation rate.

Weighted least squares (WLS) incorporates this directly by minimizing a \emph{weighted} sum of squared residuals:
\[
\sum_{i=1}^{n} w_i 
(\mu_i - \beta_0 - \bm{x}_i^{\ast\top}\bm{\beta})^2,
\qquad w_i = L_i.
\]
Thus the weights $w_i$ enter not by changing the regression equation itself, but by changing the criterion used to estimate $\bm{\beta}$. Assigning $w_i = L_i$ is approximately equivalent to weighting observations in inverse proportion to their variance, since $\mathrm{Var}(\mu_i) \propto 1/L_i$. Genes with larger callable length therefore contribute more to estimation because mutation counts are aggregated over more sites, yielding more precise estimates of the per-base mutation rate.

The regression model remains
\[
\mu_i = \beta_0 + \bm{x}_i^{\ast\top}\bm{\beta} + \varepsilon_i,
\]
and coefficients retain the same interpretation as in OLS. The weighting affects efficiency and stability of estimation, not the substantive meaning of parameters. In effect, WLS aligns the estimation procedure with the expected sampling precision of each gene’s mutation-rate estimate.

\subsubsection{Model 3: Poisson Regression for Mutation counts with a length offset} \label{subsubsec:poisson}
Models 1--2 treat the observed rate $\mu_i=C_i/L_i$ as the response. A more direct approach is to model the underlying \emph{counts} $C_i$ themselves. This avoids dividing by $L_i$ and instead incorporates gene length explicitly, reflecting the fact that mutation counts scale with the number of callable sites.

For gene $i$, let $C_i$ be the number of observed de novo mutations and let $L_i$ be its callable length in base pairs. We assume
\begin{equation}
C_i \sim \mathrm{Poisson}(\Lambda_i),
\qquad
\log \Lambda_i \;=\; \log L_i \;+\; \beta_0 \;+\; \bm{x}_i^{\ast\top}\bm{\beta},
\label{eq:poisson}
\end{equation}
where $\Lambda_i = \mathbb{E}[C_i]$ is the expected mutation count for gene $i$. The term $\log L_i$ is included as an \emph{offset}, meaning it is fixed with coefficient 1 rather than estimated. This enforces the baseline scaling $\Lambda_i \propto L_i$: doubling gene length doubles the expected count, all else equal.

With this parameterization,
\[
\Lambda_i \;=\; L_i \,\exp\!\left(\beta_0 + \bm{x}_i^{\ast\top}\bm{\beta}\right),
\]
so predictors act on the \emph{per-base} rate multiplicatively. The coefficient interpretation follows directly: $\exp(\beta_j)$ is the factor by which the expected per-base mutation rate changes for a one--standard--deviation increase in predictor $j$, holding the other predictors constant (a ``rate ratio'').

We do not assume that gene-level mutation counts are exactly Poisson. Clustering, unmodeled heterogeneity, or other processes can produce extra variation relative to Poisson expectations. We use this model as a likelihood-based specification that (i) respects the discrete nature of the data and (ii) incorporates gene length exactly as exposure, providing a complementary check on the conclusions from rate-based linear models.

\subsection{Collinearity Control and Multiplicity} \label{subsec:collinearity_fdr}
Epigenomic predictors are not statistically independent. Many chromatin marks co-occur or are mutually exclusive because they reflect overlapping aspects of chromatin organization and gene regulation. In a multivariable regression, such correlation induces \emph{multicollinearity}. Although ordinary least
squares estimators remain unbiased under collinearity, the variance of $\hat\beta_j$ increases as predictors become more correlated. As a result, confidence intervals widen, statistical power decreases, and coefficient
estimates may become numerically unstable. In the extreme case where one predictor is nearly a linear combination of others, small perturbations in the data can produce large changes in estimated coefficients, including reversals of sign. In such settings, partial effects are difficult to interpret reliably.

To quantify collinearity, we computed the Variance Inflation Factor (VIF) for each predictor. The VIF for predictor $j$ is defined as
\[
\mathrm{VIF}_j = \frac{1}{1 - R_j^2},
\]
where $R_j^2$ is obtained by regressing predictor $j$ on all remaining predictors. $\mathrm{VIF}_j$ measures how much the variance of $\hat\beta_j$ is inflated relative to the hypothetical case in which predictors are orthogonal. Large values indicate substantial redundancy. We iteratively removed predictors with $\mathrm{VIF} > 10$, a conventional threshold for severe collinearity. This step does not imply biological independence among chromatin marks; rather, it ensures that estimated partial associations are statistically identifiable and numerically stable.

Because multiple regression coefficients are evaluated simultaneously, we control the False Discovery Rate (FDR) by using the Benjamini--Hochberg procedure \cite{ControllingFalseDiscovery}. Let $m$ denote the number of tested hypotheses and $p_{(1)} \le \dots \le p_{(m)}$ the ordered $p$-values. The procedure selects the largest $k$ such that
\[
p_{(k)} \le \frac{k}{m}\alpha,
\]
and declares the first $k$ hypotheses significant. The resulting adjusted values ($q$-values) control the \emph{false discovery rate} (FDR), defined as the expected proportion of false rejections among all rejected hypotheses. FDR control is appropriate here because the objective is not to eliminate the probability of any false positive (as in family-wise error control), but to limit the expected fraction of spurious associations among the set of reported predictors. In a setting with correlated genomic features and multiple simultaneous tests, this provides a balance between discovery and error control.

\subsection{Defining a Cross-Model Predictor Set} \label{subsec:stability}
OLS, weighted least squares, and Poisson regression impose different assumptions about error structure and scale. Because no single specification can be assumed correct for sparse gene-level mutation data, we identify predictors based on consistency across models rather than on the $p$-values from any one analysis.

For each predictor $j$, we obtain a Benjamini--Hochberg adjusted $q$-value from each of the three models testing the null hypothesis $H_0:\beta_j = 0$. This yields three $q$-values per predictor, one from each specification. We then compute the arithmetic mean of these three $q$-values and use this quantity as a cross-model ranking statistic. Predictors with small mean $q$-values are those that tend to show evidence of association under multiple modeling assumptions.

We retain the five predictors with the smallest mean $q$-values. The cap of five predictors is specified in advance to promote parsimony and to limit residual collinearity in subsequent interpretation. This rule is deliberately conservative: it favors associations that are stable across reasonable statistical formulations and downweights effects that appear only under a single set of assumptions.

The stability criterion is not a claim of causality. It is a pragmatic method for prioritizing a small set of predictors in a correlated, moderately high-dimensional setting, with emphasis on robustness rather than model-specific significance.

\subsection{Interpretation of Regression Coefficients} \label{subsec:interpretation}
In all multivariable models, coefficients represent conditional (partial) associations. Each $\beta_j$ quantifies the expected change in the response associated with a one–standard–deviation increase in predictor $j$, holding all other included predictors constant; it does not measure the total or isolated effect of that feature.  Because chromatin marks and other genomic features are correlated and may both influence and reflect transcription, repair activity, and related processes, estimated coefficients should not be interpreted as direct mechanistic effects. Rather, they summarize structured statistical covariation in the observed data. Our objective is therefore limited to testing whether mutation rate—specifically its mean and dispersion—is systematically associated with epigenomic state within a functionally constrained class of genes, without inferring causal direction.

\section{Results} \label{sec:results}
We first identify which features of chromatin state and sequence composition are associated with the \emph{mean} per–base mutation rate in essential and lethal genes. Second, we test whether these same features are associated with between-gene variance and higher moments of the mutation-rate distribution. Third, we evaluate whether a population-genetic summary statistic, Tajima’s $D$, covaries with these mutation-rate properties after conditioning on chromatin predictors.

Our criterion throughout is stability of partial associations across model classes. We report effects that are consistent under OLS, weighted least squares, and count-based regression. We do not interpret coefficients as direct mechanistic effects of chromatin marks on mutation, nor do we claim that selection directly determines local mutation rates. The results are presented as structured statistical associations within a constrained gene set.

\subsection{Mean Effects: Active Chromatin Marks Predict Lower Mutation Rates} \label{subsec:mean_effects}
After VIF screening and cross-model false discovery rate control, the predictors retained for interpretation were three histone marks associated with transcriptionally active chromatin and two sequence-level variables:
\[
\{\text{H3K4me1},\ \text{H3K4me3},\ \text{H3K36ac};\ \text{GC\%},\ \text{CG methylation}\}.
\]
All predictors were standardized. Coefficients therefore represent the expected change in per–base mutation rate associated with a one–standard–deviation increase in the predictor, conditional on the remaining four predictors being fixed at their sample means (0 on the standardized scale).

In the joint OLS model with HC3 inference, the estimated partial effects were:
\[
\widehat\beta_{\text{H3K4me1}} = -1.98\times 10^{-6},
\qquad q_{\mathrm{BH}} = 9.2\times10^{-9},
\]
\[
\widehat\beta_{\text{H3K4me3}} = -9.99\times 10^{-7},
\qquad q_{\mathrm{BH}} \approx 1.1\times10^{-3},
\]
\[
\widehat\beta_{\text{H3K36ac}} = -5.74\times 10^{-7},
\qquad q_{\mathrm{BH}} \approx 1.1\times10^{-3}.
\]

All three active-chromatin marks show negative partial associations with mutation rate. In contrast, GC content had a small positive point estimate and CG methylation a small negative point estimate, but neither remained significant after FDR correction in the multivariable model. Crucially, the signs and relative magnitudes of coefficients were consistent across all three model classes (OLS with robust inference, length-weighted least squares, and Poisson regression with a length offset). 

\begin{figure}[h!]
    \centering
    \includegraphics[width=0.85\textwidth]{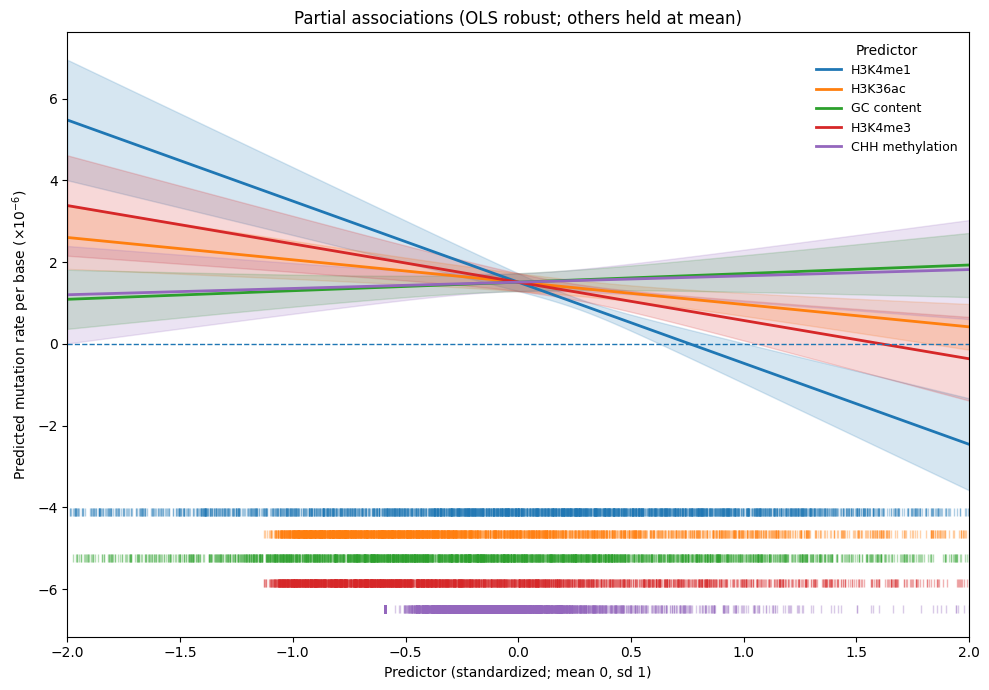}
 \caption{\textbf{Partial associations for the stable predictor set (mean model).}
Partial-effect curves from the multivariable OLS model with HC3 robust inference, restricted to essential/lethal genes. Each curve shows the expected per--base mutation rate as a single standardized predictor varies within $[-2,2]$ standard deviations, while all other predictors are fixed at their standardized means (0). Shaded regions denote 95\% heteroskedasticity-robust confidence intervals. Rug marks along the horizontal axis indicate the empirical distribution of each predictor within the plotted range; denser rugs correspond to regions with more genes and therefore stronger empirical support.}
    \label{fig:regression_lines}
\end{figure}

Figure~\ref{fig:regression_lines} displays partial-effect curves from the joint OLS model. For each predictor, values are varied within a symmetric standardized range of $[-2,2]$ while the remaining predictors are fixed at their means (0). The fitted line therefore represents the conditional expectation
\[
\widehat{\mu}_i = \widehat{\beta}_0 + \widehat{\beta}_j x_j,
\]
holding other covariates constant. Shaded bands denote 95\% confidence intervals based on HC3 robust standard errors \cite{long_using_2000}.

Rug marks along the horizontal axis show the empirical distribution of each standardized predictor within the plotted range. Dense regions of rugs indicate where most genes lie; sparse regions indicate extrapolation toward the tails. Thus the rugs identify the portion of the predictor space that is well supported by the data and distinguish it from visually extended but data-sparse regions of the fitted line.

The three histone marks exhibit monotone negative partial slopes across the region containing most observations. In contrast, GC content and CG methylation show shallow slopes centered near zero, and their confidence bands overlap zero across much of the empirical range once chromatin marks are held constant.

These results are statistically consistent with prior observations that transcriptionally active chromatin is associated with lower mutation rates \cite{monroe_mutation_2022}. They do not establish mechanism. The coefficients quantify conditional associations within a constrained gene set. Restricting the analysis to essential and lethal genes reduces variation in both predictors and mutation outcomes, which may attenuate detectable sequence-level effects. In addition, despite VIF-based pruning, histone marks remain correlated. The reported coefficients therefore reflect partial associations within a correlated system rather than independent causal contributions.

\subsection{Variance and Higher Moments: Variance Suppression} \label{subsec:moments}
Mean effects describe how predictors covary with the expected per--base mutation rate, but they do not describe how mutation rates are distributed across genes. If chromatin context influences repair efficiency, it should affect not only the mean rate but also the \emph{between-gene variability} of rates. We therefore examined variation beyond the mean by modeling the conditional variance and, secondarily, higher moments.

To assess variance structure, we first fit the mean model and computed residuals $\hat{\varepsilon}_i$ from Eq.~\eqref{eq:ols}. We then modeled the squared residuals as a function of the same standardized predictors using a log-variance specification. Coefficients are reported as the percent change in residual variance associated with a one--standard--deviation increase in a predictor, conditional on the other predictors.

The dominant pattern is strong variance suppression for marks of transcriptionally active chromatin:
$$\text{H3K4me1: } -84.9\% \quad (q_{\mathrm{BH}}
\ll 10^{-10}),$$
$$\text{H3K4me3: } -58.9\% \quad (q_{\mathrm{BH}} 
\ll 10^{-10}),$$
$$\text{H3K36ac: } -36.2\% \quad (q_{\mathrm{BH}} 
\ll 10^{-10}).$$

Thus, within essential/lethal genes, higher levels of these marks are associated not only with lower expected mutation rates, but with substantially reduced heterogeneity in rates across loci.

\begin{figure}[h!]
    \centering
    \includegraphics[width=0.7\textwidth]{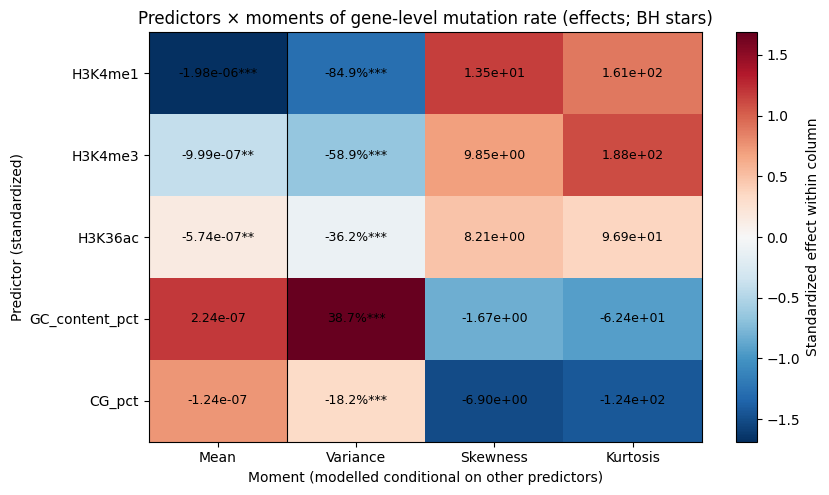}
    \caption{\textbf{Predictors $\times$ moments of the per--base mutation rate (essential/lethal genes).}
Each cell reports the estimated regression coefficient for a given predictor (row) and moment (column). ``Mean'' shows partial effects on the expected per--base mutation rate from the multivariable OLS model. ``Variance'' reports the percent change in residual variance per one--standard--deviation increase in the predictor, based on a log-variance regression of $\log(\hat{\varepsilon}_i^2)$ on the same predictors. ``Skewness'' and ``Kurtosis'' report coefficients from regressions on standardized-residual moments ($u_i^3$ and $u_i^4-3$, respectively), where $u_i$ denotes residuals scaled by the fitted variance and clipped at the 0.5th--99.5th percentiles to limit outlier leverage. 
Colors represent column-wise standardized effect sizes (z-scores; for visualization only), while annotations show raw effect estimates. Stars denote Benjamini--Hochberg FDR significance within each moment. The strongest and most consistent pattern is variance suppression associated with active-chromatin marks (H3K4me1, H3K4me3, H3K36ac); effects on skewness and kurtosis are not statistically supported.}

    \label{fig:moment_heatmap}
\end{figure}

Sequence features show a different pattern. GC content is associated with increased variance,
\[
\text{GC content: } +38.7\% \quad (q_{\mathrm{BH}} < 10^{-11}),
\]
whereas CG methylation shows a modest stabilizing association,
\[
\text{CG methylation: } -18.2\% \quad (q_{\mathrm{BH}} \approx 5\times10^{-6}).
\]
We also estimated models for skewness and kurtosis, but found no statistically supported associations after multiple-testing correction. Figure~\ref{fig:moment_heatmap} summarizes the mean and moment regressions. The main empirical result is not a detectable change in tail shape, but a pronounced reduction in variance associated with active-chromatin marks. In this gene set, the predictors most strongly associated with lower mean mutation rate are associated even more strongly with reduced between-gene fluctuation in mutation rate.

\subsection{Association with Tajima's $D$: Suggestive but Underpowered} \label{subsec:tajima}
We use Tajima’s $D$ to ask whether variation in long-term constraint within the essential/lethal gene set covaries with epigenomic state and with the distribution of mutation rates across genes. Because Tajima’s $D$ is shaped by demography and linked selection as well as purifying selection, we treat it as an imperfect proxy rather than a direct measure of constraint intensity. For comparability across models, we define a standardized score
\[
S_i \;=\; -\,\frac{D_i-\bar D}{\mathrm{SD}(D)},
\]
so that larger $S_i$ corresponds to more negative Tajima’s $D$.

We first asked whether $S$ covaries with the predictor set. For each predictor $X_j^\ast$, we fit a multivariable regression with $S$ as an additional covariate. The coefficient for $S$ then represents the expected change (in standard-deviation units) in predictor $j$ per one--standard--deviation increase in $S$, conditional on the remaining predictors. Only GC content showed a statistically supported association:
\[
\hat\gamma_{\mathrm{GC}} = 0.048,\quad 95\%\,\mathrm{CI}=[0.015,\,0.081],\quad q_{\mathrm{BH}}=0.020.
\]
Partial slopes for H3K4me1, H3K4me3, H3K36ac, and CG methylation were small and not significant after FDR correction. Rank correlations between $S$ and individual predictors were near zero, implying that the GC signal appears primarily as a conditional association once correlated chromatin marks are accounted for. We therefore interpret it as a partial statistical relationship, not as an isolated mechanistic link.

\begin{figure}[h!]
  \centering
  \includegraphics[width=\textwidth]{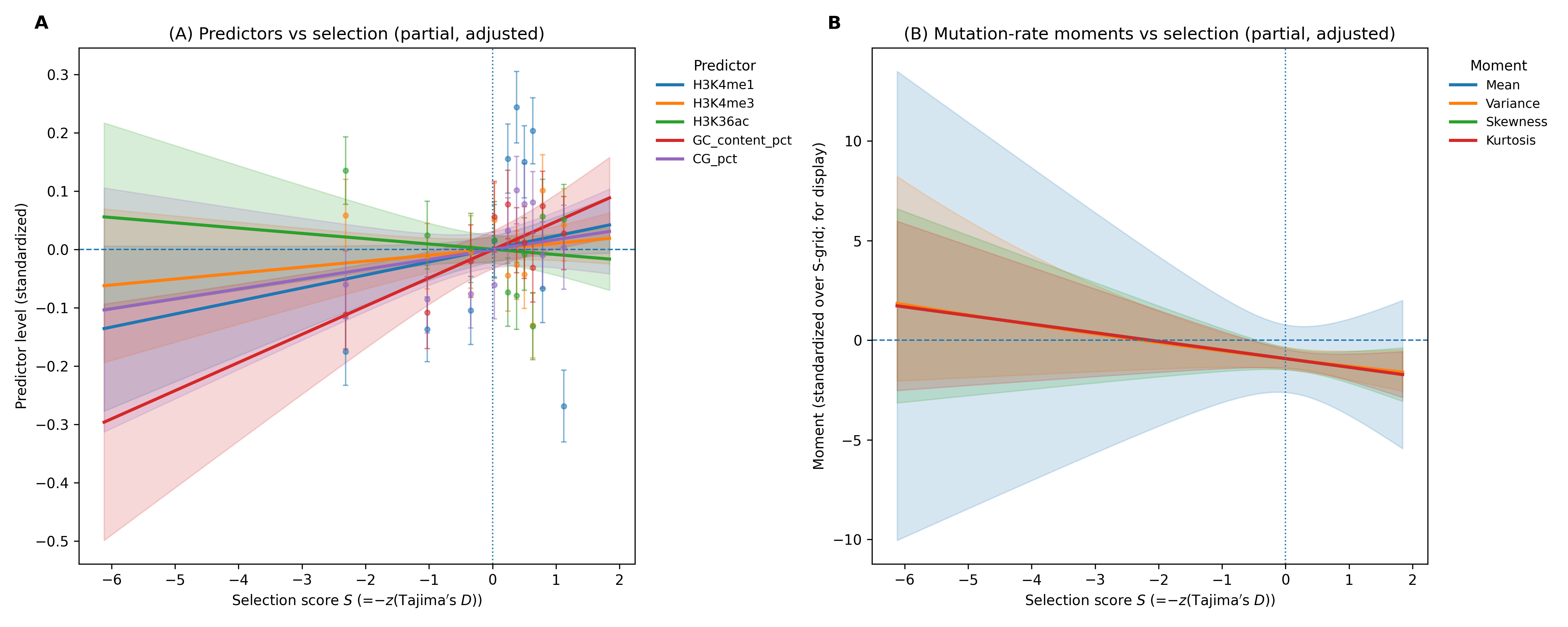}
  \caption{\textbf{Selection proxy versus predictors and mutation-rate moments in essential/lethal genes.} Selection intensity is indexed by $S = -z(\mathrm{Tajima's}\ D)$, so larger $S$ corresponds to more negative Tajima’s $D$ (i.e., stronger inferred purifying selection). The sign reversal is for interpretive clarity only. 
\emph{(A)} Partial associations between each standardized predictor and $S$, from regressions $X_j^\ast \sim S + X_{-j}^\ast$ with heteroskedasticity-robust 95\% confidence intervals. Lines show adjusted predictions (others fixed at 0); points are binned means $\pm$\,SEM. 
\emph{(B)} Predicted mean and higher-moment summaries from models including $S$ and chromatin predictors (others fixed at 0). Curves are standardized over the plotted $S$ range for display only. Shaded regions denote robust 95\% confidence intervals.}
  \label{fig:tajimas_trends}
\end{figure}

We next asked whether $S$ predicts mutation-rate properties after conditioning on epigenomic context. We added $S$ to the mean and moment models and estimated its partial association with the per--base mutation rate and its conditional variance, skewness, and kurtosis. All estimated slopes were negative—consistent with lower mean and reduced variability at more negative Tajima’s $D$—but none were statistically supported after multiple-testing correction:
\[
\text{mean: }\hat\beta_S=-5.6\times10^{-8}\ (q_{\mathrm{BH}}=0.65),\qquad
\text{variance: }-5.2\%\ (q_{\mathrm{BH}}=0.19),
\]
\[
\text{skewness: }-13.7\ (q_{\mathrm{BH}}=0.27),\qquad
\text{kurtosis: }-276\ (q_{\mathrm{BH}}=0.20).
\]
Figure~\ref{fig:tajimas_trends} summarizes how the standardized selection proxy  $S=-z(\mathrm{Tajima's}\ D)$ relates to epigenomic predictors and to mutation-rate properties within essential/lethal genes. (Standardization centers Tajima’s $D$ and rescales it to unit variance; the negative sign ensures that larger $S$ corresponds to more negative $D$.) 

Panel A displays partial associations between each standardized predictor and $S$, estimated from regressions of the form  $X_j^\ast \sim S + X_{-j}^\ast$.  The solid lines are adjusted predictions with the remaining predictors fixed at their standardized means (0).  Points represent binned empirical means of the predictor across quantiles of $S$, with vertical bars denoting $\pm$\,SEM; they provide a descriptive summary of the observed data and are not model-based fits. 

Panel B shows predicted mean and higher-moment summaries from models that include $S$ in addition to the chromatin predictors, again holding predictors at their means. For display only, the moment curves are standardized across the plotted $S$ range so that effect magnitudes can be visually compared; this rescaling does not alter slope direction or statistical inference. 

Across moments, the estimated slopes are directionally consistent with lower mean and reduced variability at more negative Tajima’s $D$, but uncertainty intervals are wide and none of the $S$ effects meet the false discovery rate threshold. Within this restricted gene set, the data provide limited evidence that variation in Tajima’s $D$ explains mutation-rate structure beyond that already captured by chromatin context.

\section{Discussion} \label{sec:discussion}
Within essential and lethal genes of \textit{Arabidopsis thaliana}, mutation rate is not uniform across loci. In multivariable models that account for gene length and correlated predictors, three marks of transcriptionally active chromatin (H3K4me1, H3K4me3, H3K36ac) are jointly associated with lower mean per--base mutation rate. The stronger signal, however, is in the second moment: the same marks are associated with large reductions in between-gene variance.  The pattern is stable across alternative specifications that differ in error structure and weighting, suggesting that they do not arise from a particular modeling choice.

The appropriate inference is structural. Histone modifications are not interpreted as direct causal suppressors of mutation. They index chromatin environments that differ in transcriptional activity, accessibility, and repair dynamics. The result is therefore not that a specific mark ``reduces mutation,''but that mutational input varies systematically with the genomic context in which a gene is embedded.

The variance effect is central because it separates two quantities that are often conflated: the expected mutation rate and the variance of mutation risk across loci. Standard treatments summarize mutation by a single parameter $\mu$. Our results show that, even within a restricted set of strongly constrained genes, the distribution of mutation rates is nontrivial and systematically patterned. Active chromatin is associated not only with lower expected mutation rates but with reduced locus-to-locus fluctuation. By contrast, GC content contributes primarily to increased variance once chromatin features are held constant. In this gene set, sequence composition appears to modulate volatility more than baseline intensity.

This distinction aligns with recent theoretical work in which mutation rate is allowed to vary stochastically rather than remain constant. In models where mutation rate varies across generations—for example in response to stress or environmental conditions—the evolutionary consequences depend not only on the mean mutation rate but on the distribution of the rate \cite{heinrich-moraEvolutionStochasticTransmission2026}.  Our results address an analogous problem across genomic space: loci differ not only in their mean mutational input but in the stability of that input, and chromatin context is one axis along which that stability varies.

Tajima’s $D$ provides a limited and indirect index of constraint within essential genes. After conditioning on chromatin predictors, estimated associations between $S=-z(D)$ and the mean and higher moments of mutation rate distribution are directionally consistent with stronger constraint corresponding to lower mean and reduced variance, but are statistically imprecise. Given the restricted polymorphism in essential genes and the demographic sensitivity of Tajima’s $D$, this lack of precision reflects the limits of the data rather than contradicting the broader structural pattern.

Two implications follow. Empirically, the association between active chromatin and reduced mutational variability should be tested across taxa and functional gene classes, and in conjunction with environmentally induced changes in mutation rate. Theoretically, models that treat mutation rate as constant may remain useful approximations at coarse scales, but they omit heterogeneity that is evident at the level of individual loci. A more complete account of genome evolution requires attention not only to the genome-wide mean mutation rate, but to how mutational input is distributed—and stabilized—across genomic architecture.

In summary, mutation rate is neither a purely exogenous constant nor an unconstrained random fluctuation. Our results suggests that the distribution of mutation rate is a structured quantity whose distribution is shaped by genomic organization and, in other contexts, by physiological and environmental states. In the present data, active chromatin is associated with both reduced mutational input and reduced variability among constrained loci. If fitness effects are nonlinear—so that rare mutational bursts in essential genes carry disproportionate cost—then the distribution of mutation rates across loci becomes evolutionarily relevant. Under such conditions, selection may favor genomic configurations that incidentally stabilize mutational input in sensitive regions, even without altering the genome-wide mean. Any theory that treats mutation solely through its genome-wide mean will miss this layer of structure.

\newpage
\section*{Author Contributions}

\textbf{Elisa Heinrich-Mora:} Conceptualization; Data curation; Formal analysis; Investigation; Methodology; Visualization; Writing – original draft; Writing – review $\&$ editing.

\noindent \textbf{Marcus W. Feldman:} Conceptualization; Methodology; Supervision; Validation; Writing – review $\&$ editing; Resources; Funding acquisition.

\section*{Conflict of Interest} 
The authors declare that they have no conflict of interest.

\section*{Data and Code Availability} 
The datasets analyzed during the current study were derived from publicly available sources reported in Monroe et al. \cite{monroe_mutation_2022}. Processed datasets used in the present analysis are available in the accompanying code repository.

The complete codebase used to perform the analyses and generate the figures is publicly available at:
\begin{center}
\url{https://github.com/elisaheinrichmora/Evo_Mutation_Distribution_Athaliana} 
\end{center}

\section*{Funding}
This research was supported in part by the Center for Computational, Evolutionary and Human Genomics (CEHG) at Stanford University.

\bibliographystyle{abbrv}
\bibliography{references}

\end{document}